\documentclass[showkeys,showpacs,prd,aps,floats,eqsecnum,preprint]{revtex4}

\begin{document}

\title{Regularizing role of teleparallelism}

\author{Tiago Gribl Lucas}
\email{gribl@ift.unesp.br}
\affiliation{Instituto de F\'{\i}sica Te\'orica, 
UNESP-Universidade Estadual Paulista,  
Rua Dr.\ Bento Teobaldo Ferraz 271, 01140-070 S\~ao Paulo, Brazil}
\author{Yuri N.~Obukhov}
\email{yo@ift.unesp.br}
\affiliation{Instituto de F\'{\i}sica Te\'orica, 
UNESP-Universidade Estadual Paulista,  
Rua Dr.\ Bento Teobaldo Ferraz 271, 01140-070 S\~ao Paulo, Brazil}
\affiliation{Department of Theoretical Physics, 
Moscow State University, 117234 Moscow, Russia}
\author{J.G. Pereira}
\email{jpereira@ift.unesp.br}
\affiliation{ Instituto de F\'{\i}sica Te\'orica, 
UNESP-Universidade Estadual Paulista,
Rua Dr.\ Bento Teobaldo Ferraz 271, 01140-070 S\~ao Paulo, Brazil}

\begin{abstract}
The properties of the gravitational energy-momentum 3-form and of
the superpotential 2-form are discussed in the covariant teleparallel 
framework, where the Weitzenb\"ock connection represents inertial 
effects related to the choice of the frame. Due to its odd asymptotic 
behavior, the contribution of the inertial effects often yields 
unphysical (divergent or trivial) results for the total energy of 
the system. However, in the covariant teleparallel approach, the 
energy is always finite and nontrivial. The teleparallel connection 
plays a role of a regularizing tool which subtracts the inertial 
effects without distorting the true gravitational contribution. As 
a crucial test of the covariant formalism, we reanalyze the computation 
of the total energy of the Schwarzschild and the Kerr solutions.
\end{abstract}



\maketitle

\section{Introduction}

The problem of defining an energy-momentum density for the gravitational
field belongs to the oldest in modern theoretical physics. The concepts 
of energy and momentum are fundamental ones in classical field theory. 
Within the general Lagrange-Noether approach, conserved currents arise 
from the invariance of the classical action under transformations of 
fields and spacetime variables. In particular, energy and momentum are 
related to time and space translations. However, due to the geometric 
nature of the gravitational theory and because of the equivalence principle, 
which identifies locally gravitation and inertia, the definition of 
gravitational energy remained unsolved for many years.
In general, there are no symmetries in Riemannian manifolds that can
be used to generate the corresponding conserved energy-momentum currents.
It is possible, though, to associate energy and momentum to 
asymptotically flat gravitational field configurations. The history of
the problem and some of the corresponding achievements is described 
in reviews \cite{Trautman,Faddeev,Szabados,Blag,Nester}, for example. 

On the other hand, although equivalent to general relativity, the gauge 
structure of teleparallel gravity gives rise to several conceptual and 
practical differences in relation to the geometric structure of general 
relativity. An important difference is that it is possible to distinguish 
gravitation and inertia \cite{Einstein05}. Since inertia is in the realm 
of the pseudotensor behavior of the usual expressions for the gravitational 
energy-momentum density, it turns out possible in teleparallel gravity 
to write down a tensorial expression for such density \cite{PeLu1}. With 
the purpose of getting a deeper insight into the covariant teleparallel
formalism, as well as to test how it works, we reanalyze the computation 
of the total energy of the two important examples, namely, the Schwarzschild 
and Kerr solutions.

The paper is organized as follows. Using the language of exterior 
forms, we give in Sec.~\ref{TG} an outline of the teleparallel 
approach to gravity. In Sec.~\ref{EMC} we present the covariant 
formalism for the gravitational energy-momentum. In simple terms, 
it means that a general relativistic system is described not by a 
single variable $\vartheta^\alpha$, as was done in the pure tetrad
approach \cite{Moller,Pele,Kaempfer,Rodicev,Cho,HSh79,Meyer}, but 
by the pair $(\vartheta^\alpha, \Gamma_\alpha{}^\beta)$. 
The tetrad $\vartheta^\alpha$ is responsible for the gravitational
effects, but its form also reflects the choice of the reference system.
This inevitably brings in the inertial phenomena which are mixed up with
the truly gravitational effects. The introduction of the teleparallel 
connection $\Gamma_\alpha{}^\beta$ makes it possible to deal with the
inertial effects in a constructive way. Specifically, we demonstrate
in Sec.~\ref{ES} that, due to an inconvenient choice of a reference
system, the traditional computation of the total energy of the 
Schwarzschild solution can yield either a divergent or a vanishing 
result. With an account of the teleparallel connection, we can circumvent 
such results. In our covariant formalism, the Weitzenb\"ock connection 
acts as a regularizing tool that separates the inertial contribution 
and provides the physically meaningful result for all reference frames.
The notion of a proper tetrad, introduced in Sec.~\ref{ES2}, plays
a central role in this approach. The results obtained 
are then further generalized to the case of the Kerr solution in
Sec.~\ref{KS}. Finally, in Sec.~\ref{DC} we summarize our results,
namely, that the covariant teleparallel formalism automatically
regularizes the computations, always yielding the physically relevant
solution. 

Our general notations are as in \cite{HMMN95}. In particular, we use the 
Latin indices $i,j,\dots$ for local holonomic spacetime coordinates and the
Greek indices $\alpha,\beta,\dots$ label (co)frame components. Particular 
frame components are denoted by hats, $\hat 0, \hat 1$, etc. As usual,
the exterior product is denoted by $\wedge$, while the interior product of a 
vector $\xi$ and a $p$-form $\Psi$ is denoted by $\xi\rfloor\Psi$. The vector 
basis dual to the frame 1-forms $\vartheta^\alpha$ is denoted by $e_\alpha$ 
and they satisfy $e_\alpha\rfloor\vartheta^\beta=\delta^\beta_\alpha$. 
Using local coordinates $x^i$, we have $\vartheta^\alpha=h^\alpha_idx^i$ and 
$e_\alpha=h^i_\alpha\partial_i$. We define the volume 4-form by 
$\eta:=\vartheta^{\hat{0}}\wedge\vartheta^{\hat{1}}\wedge
\vartheta^{\hat{2}}\wedge\vartheta^{\hat{3}}$. Furthermore, with the help of 
the interior product we define $\eta_{\alpha}:=e_\alpha\rfloor\eta$, 
$\eta_{\alpha\beta}:=e_\beta\rfloor\eta_\alpha$, $\eta_{\alpha\beta\gamma}:=
e_\gamma\rfloor\eta_{\alpha\beta}$ which are bases for 3-, 2- and 1-forms 
respectively. Finally, $\eta_{\alpha\beta\mu\nu} = e_\nu\rfloor\eta_{\alpha
\beta\mu}$ is the Levi-Civita tensor density. The $\eta$-forms satisfy the
useful identities:
\begin{eqnarray}
\vartheta^\beta\wedge\eta_\alpha &=& \delta_\alpha^\beta\eta ,\\
\vartheta^\beta\wedge\eta_{\mu\nu} &=& \delta^\beta_\nu\eta_{\mu} -
\delta^\beta_\mu\eta_{\nu},\label{veta1}\\ \label{veta}
\vartheta^\beta\wedge\eta_{\alpha\mu\nu}&=&\delta^\beta_\alpha\eta_{\mu\nu
} + \delta^\beta_\mu\eta_{\nu\alpha} + \delta^\beta_\nu\eta_{\alpha\mu},\\
\label{veta2}
\vartheta^\beta\wedge\eta_{\alpha\gamma\mu\nu}&=&\delta^\beta_\nu\eta_{\alpha
\gamma\mu} - \delta^\beta_\mu\eta_{\alpha\gamma\nu} + \delta^\beta_\gamma
\eta_{\alpha\mu\nu} - \delta^\beta_\alpha\eta_{\gamma\mu\nu}.
 \end{eqnarray}
The line element $ds^2 = 
g_{\alpha\beta}\vartheta^\alpha\otimes\vartheta^\beta$ is defined by the 
spacetime metric $g_{\alpha\beta}$. 

\section{Teleparallel gravity}\label{TG}

The teleparallel approach is based on the gauge theory of translations. 
Without going into the subtleties of the corresponding gauge-theoretic 
scheme (for an advanced reading, see \cite{Cho,Gron,Muench97,dea,Tres},
for example), one can view the coframe $\vartheta^\alpha = h^\alpha_idx^i$
(tetrad) as a one-form that plays the role of the gauge translational 
potential of the gravitational field. Einstein's general relativity 
theory can be reformulated as the teleparallel theory. Geometrically, 
one can view the teleparallel gravity as a special (degenerate) case 
\cite{telemag,telemag2,noninv}
of the metric-affine gravity in which the coframe $\vartheta^\alpha$ and
the local Lorentz connection $\Gamma_\alpha{}^\beta$ are subject
to the distant parallelism constraint $R_\alpha{}^\beta = 0$. 
The torsion 2-form 
\begin{equation}
T^\alpha = d\vartheta^\alpha+\Gamma_\beta{}^{\alpha}\wedge\vartheta^\beta,
\label{defT}
\end{equation}
arises as the gravitational gauge field strength, with $\Gamma_\beta{}^{\alpha}$
the Weitzenb\"ock connection. As is well known, torsion $T^\alpha$ can
be decomposed into three irreducible pieces: the tensor part, the trace, 
and the axial trace, given respectively by 
\begin{eqnarray}\label{Fi1}
{}^{(1)}T^{\alpha}&:=& T^{\alpha}-{}^{(2)}T^{\alpha}-{}^{(3)}T^{\alpha},\\
{}^{(2)}T^{\alpha}&:=& \frac{1}{3}\,\vartheta^\alpha\wedge\left(e_\beta\rfloor
T^\beta\right),\label{Fi2}\\
{}^{(3)}T^{\alpha}&:=&\frac{1}{3}\,e^\alpha\rfloor\left(\vartheta^\beta\wedge
T_\beta\right).\label{Fi3}
\end{eqnarray}

A Yang-Mills type Lagrangian is then constructed as a quadratic polynomial
in torsion. In the so-called teleparallel equivalent gravity model, 
the Lagrangian reads
\begin{equation} \label{V2}
V = -\,{\frac 1 {2\kappa}}T^{\alpha}\wedge{}^\star\left(
{}^{(1)}T_{\alpha}- 2{}^{(2)}T_{\alpha} -{1\over 2}{}^{(3)}T_{\alpha}\right),
\end{equation}
where $\kappa=8\pi G/c^3$, and ${}^\star$ denotes the Hodge dual in the metric
$g_{\alpha\beta}$. The latter is assumed to be the flat Minkowski metric 
$g_{\alpha\beta} = o_{\alpha\beta} :={\rm diag}(+1,-1,-1,-1)$, and it is 
used to raise and lower the Greek (local frame) indices.

The teleparallel field equations are obtained from the variation of
the total action with respect to the coframe:
\begin{equation}
DH_\alpha - E_\alpha = \Sigma_\alpha.\label{geq}
\end{equation}
Here $D$ denotes the covariant exterior derivative, i.e., $DH_\alpha=d
H_\alpha-\Gamma_\alpha^{\ \beta}\wedge H_\beta$. The translational momentum
and the canonical energy-momentum are, respectively: 
\begin{eqnarray}
H_{\alpha} = -\,{\frac {\partial V} {\partial T^\alpha}} &=&
\,{1\over \kappa}\,{}^\star\!\left({}^{(1)}T_{\alpha} - 2\,{}^{(2)}T_{\alpha}
- {1\over 2}\,{}^{(3)}T_{\alpha}\right),\label{Ha0}\\
E_\alpha = {\frac {\partial V} {\partial \vartheta^\alpha}} &=&
e_\alpha\rfloor V + (e_\alpha\rfloor T^\beta)\wedge H_\beta.\label{Ea0}
\end{eqnarray}
In terms of $H_{\alpha}$, the Lagrangian (\ref{V2}) is recast in the form
\begin{equation}
V = -\,{\frac 12}\,T^\alpha\wedge H_\alpha.\label{Vlag}
\end{equation}

We remark that the resulting model is degenerate from the
metric-affine viewpoint, because the variational derivatives of the 
action with the respect to the metric and connection are trivial. This 
means that the field equations are satisfied for any $\Gamma_\alpha{}^\beta$. 
However, as we are going to see, the presence of the connection field plays 
an important {\it regularizing} role. Furthermore, in its presence the 
teleparallel gravity becomes explicitly covariant under local Lorentz 
transformations of the coframe. In particular, the Lagrangian (\ref{V2}) 
is invariant under the changes
\begin{equation}\label{cofcontrans}
\vartheta'^\alpha = L^\alpha{}_{\beta}\vartheta^\beta,\qquad
\Gamma'_\alpha{}^{\beta} = (L^{-1})^\mu{}_{\alpha}\Gamma_\mu{}^\nu
L^\beta{}_{\nu} + L^\beta{}_{\gamma}d(L^{-1})^\gamma{}_{\alpha},
\end{equation}
with $L^\alpha{}_{\beta} \equiv L^\alpha{}_{\beta}(x)\in SO(1,3)$. 
In contrast to this, the Lagrangian of the pure tetrad gravity, which 
is obtained when we put $\Gamma_\alpha{}^\beta = 0$ for all frames, is 
only quasi-invariant---it changes by a total divergence.

The connection $\Gamma_\alpha{}^\beta$ can be decomposed into Riemannian 
and post-Riemannian parts as
\begin{equation}
\Gamma_\alpha{}^\beta = \tilde{\Gamma}_\alpha{}^\beta -
K_\alpha{}^\beta .\label{gagaK}
\end{equation}
Here, $\tilde{\Gamma}_\alpha{}^\beta$ is the purely Riemannian connection
and $K_\alpha{}^\beta$ is the contortion which is
related to the torsion by the identity
\begin{equation}
T^\alpha=K^\alpha{}_\beta\wedge\vartheta^\beta.\label{contor}
\end{equation}
One can then show that, due to geometric identities \cite{PGrev}, 
the translational momentum (\ref{Ha0}) can be written as
\begin{equation}
H_\alpha ={\frac 1 {2\kappa}}\,K^{\mu\nu}\wedge\eta_{\alpha\mu\nu}.\label{H0K}
\end{equation}

A crucial property of the teleparallel framework is that the Weitzenb\"ock
connection $\Gamma_\alpha{}^\beta$ actually represents inertial 
effects that arise from the choice of the reference system. Due to 
its odd asymptotically behavior, the 
inertial contributions in many cases yield unphysical results for the 
total energy of the system, producing either trivial or divergent 
answers. We will show in this paper that the computation of the 
energy in the covariant teleparallel approach always yields finite 
and physically correct results. In this sense, we can say that the
teleparallel connection acts
as a regularizing tool which helps to eliminate the inertial effects
without distorting the true gravitational contribution. 

It is worthwhile to mention that the Lagrangian (\ref{Vlag}) differs 
from the Hilbert-Einstein Lagrangian by a total derivative (surface
term). Correspondingly, the field equation (\ref{geq}) coincide with
Einstein's gravitational field equation. In this sense, the physical
contents of the two theories is the same.

\section{Energy-momentum conservation}\label{EMC}

We begin by rewriting the field equation (\ref{geq}) in the Maxwell-type
form:
\begin{equation}
DH_\alpha = E_\alpha + \Sigma_\alpha.\label{DH0}
\end{equation}
{}The analogy with the electromagnetism is obvious. The Maxwell 2-form
$F=dA$ represents the gauge field strength of the electromagnetic
potential 1-form $A$. From the Lagrangian $V(F)$, the 2-form of the
electromagnetic excitations is defined by $H = - \partial V/\partial
F$, and the field equation reads $dH = J$, where $J$ is the 3-form of
the electric current density of matter. In view of the nilpotency of
the exterior differential, $dd \equiv 0$, the Maxwell equation yields
the conservation law of the electric current, $dJ = 0$. 

In contrast to electrodynamics, gravity is a self-interacting field, and
the gauge field potential 1-form $\vartheta^\alpha$ carries an ``internal'' 
index $\alpha$. The gauge field strength 2-form $T^\alpha = D\vartheta^\alpha$
is now defined by the covariant derivative of the potential (compare
with $F=dA$). The gravitational field excitation 2-form $H_\alpha$ is
introduced by (\ref{Ha0}), in a direct analogy to the Maxwell theory
(recall $H = - \partial V/\partial F$). Finally, we observe that as
compared to the Maxwell field equation $dH = J$, the gravitational
field equation (\ref{DH0}) contains now the covariant derivative $D$, 
and in addition, the right-hand side is represented by a modified current
3-form, $E_\alpha + \Sigma_\alpha$. The last term is the energy-momentum
of matter, and we naturally conclude that the 3-form $E_\alpha$ describes
the {\it energy-momentum current of the gravitational field}. Its 
presence in the right-hand side of the field equation (\ref{DH0}) 
reflects the self-interacting nature of the gravitational field, and
such contribution is absent in the linear electromagnetic theory.

We can complete the analogy with electrodynamics by deriving the 
corresponding conservation law. Indeed, since $DD \equiv 0$ for the 
teleparallel connection, (\ref{DH0}) tells us that the sum of the 
energy-momentum currents of gravity and matter, $E_\alpha + \Sigma_\alpha$, 
is covariantly conserved \cite{PeLu1}, 
\begin{equation}
D(E_\alpha + \Sigma_\alpha) = 0.\label{DE1}
\end{equation}
This law is consistent with the covariant transformation properties 
of the currents $E_\alpha$ and $\Sigma_\alpha$.

One can rewrite the conservation of energy-momentum in terms of the
ordinary derivatives. Using the explicit expression $DH_\alpha = dH_\alpha
- \Gamma_\alpha{}^\beta\wedge H_\beta$, the field equation (\ref{geq}) 
and (\ref{DH0}) can be recast in an alternative form
\begin{equation}
dH_\alpha = {\cal E}_\alpha + \Sigma_\alpha,\label{DH1}
\end{equation}
where ${\cal E}_\alpha = E_\alpha + \Gamma_\alpha{}^\beta\wedge H_\beta$. 
Accordingly, (\ref{DH1}) yields a usual conservation law with the 
ordinary derivative
\begin{equation}
d({\cal E}_\alpha + \Sigma_\alpha) = 0.\label{DE2}
\end{equation}

The 3-form $E_\alpha$ describes the gravitational energy-momentum in a
covariant way, whereas the 3-form ${\cal E}_\alpha$ is a non-covariant
object. In terms of components, it gives rise to the energy-momentum
pseudotensor. It is worthwhile to note that $H_\alpha$ plays the 
role of energy-momentum superpotential both for the
covariant energy-momentum current $(E_\alpha + \Sigma_\alpha)$ and
for the total (including inertia) non-covariant current
$({\cal E}_\alpha + \Sigma_\alpha)$. 

The $\eta$-forms (defined above) serve as the basis of the spaces of forms 
of different rank, and when we expand the above objects with respect to
the $\eta$-forms, the usual tensor formulation is recovered. Explicitly, 
\begin{equation}
H_\alpha = {\frac 1\kappa}\,S_\alpha{}^{\mu\nu}\,\eta_{\mu\nu}.\label{Ha}
\end{equation}
Here $S_\alpha{}^{\mu\nu} = - S_\alpha{}^{\nu\mu}$ is constructed from the
contortion tensor in a usual way \cite{telemag}. 

Analogously, we have explicitly for the gravitational energy-momentum
\begin{equation}
E_\alpha = {\frac 12}\left[(e_\alpha\rfloor T^\beta)\wedge H_\beta 
- T^\beta\wedge(e_\alpha\rfloor H_\beta)\right].\label{Ea00}
\end{equation}
Substituting here (\ref{Ha}) and $T^\alpha = {\frac 12}\,T_{\mu\nu}{}^\alpha
\,\vartheta^\mu\wedge\vartheta^\nu$, and using (\ref{veta1})-(\ref{veta2}), 
we find 
\begin{equation}
E_\alpha = t_\alpha{}^\beta\,\eta_\beta,\qquad t_\alpha{}^\beta = 
{\frac 1{2\kappa}}\left(4T_{\alpha\nu}{}^\lambda S_\lambda{}^{\beta\nu}
- T_{\mu\nu}{}^\lambda S_\lambda{}^{\mu\nu}\,\delta_\alpha^\beta
\right).\label{Ea1}
\end{equation}
Similarly, we have
\begin{equation}
{\cal E}_\alpha = j_\alpha{}^\beta\,\eta_\beta,\qquad j_\alpha{}^\beta = 
{\frac 1{2\kappa}}\left(4T_{\alpha\nu}{}^\lambda S_\lambda{}^{\beta\nu}
- T_{\mu\nu}{}^\lambda S_\lambda{}^{\mu\nu}\,\delta_\alpha^\beta
+ 4\Gamma_{\nu\alpha}{}^\lambda\,S_\lambda{}^{\beta\nu}\right).\label{Ea2}
\end{equation}
Now, whereas $t_\alpha{}^\beta$ is a true tensor, because it depends 
explicitly on the Weitzenb\"ock connection $\Gamma_{\nu\alpha}{}^\lambda$,
the current $j_\alpha{}^\beta$ is a pseudotensor. Since the Weitzenb\"ock
connection $\Gamma_{\nu\alpha}{}^\lambda$ represents the inertial effects 
related to the choice of the frame, we see clearly that the origin of 
the pseudotensor behavior of the usual energy-momentum densities is 
that they include those inertial effects \cite{PeLu1}.

Taking into account the analogous expansion of the matter energy-momentum,
$\Sigma_\alpha = \Sigma_\alpha{}^\beta\,\eta_\beta$, which introduces the 
energy-momentum tensor $\Sigma_\alpha{}^\beta$, and using (\ref{Ha}) 
and (\ref{Ea1}), we easily recover the field equation in tensor language 
(used, for example, in \cite{dAGP00}). Note that the conservation laws 
(\ref{DE1}) and (\ref{DE2}) coincide when we put $\Gamma_\alpha{}^\beta 
= 0$. The last term in (\ref{Ea2}) then disappears, whereas torsion 
reduces to the anholonomity 2-form, $T^\alpha = F^\alpha = d\vartheta^\alpha$. 
We denote the corresponding energy-momentum and superpotential 
with a tilde: 
\begin{equation}
\widetilde{E}_\alpha
= E_\alpha\vline_{\Gamma_\alpha{}^\beta = 0},\qquad \widetilde{H}_\alpha
= H_\alpha\vline_{\Gamma_\alpha{}^\beta = 0}.\label{HEtilde}
\end{equation}
The properties of these
quantities and their use for the computation of the total energy of the
exact solutions was discussed in \cite{conserved,noninv}. Explicitly,
we have
\begin{eqnarray}
\widetilde{H}_\alpha &=& {\frac 1 {2\kappa}}\,\tilde{\Gamma}^{\beta\gamma}
\wedge\eta_{\alpha\beta\gamma},\label{Ht}\\
\widetilde{E}_\alpha &=& {\frac 12}\left[(e_\alpha\rfloor d\vartheta^\beta)
\wedge \widetilde{H}_\beta - d\vartheta^\beta\wedge(e_\alpha\rfloor 
\widetilde{H}_\beta)\right].\label{Et}
\end{eqnarray}

\section{Energy of the Schwarzschild solution}\label{ES}

In this section we will demonstrate the regularizing role of the 
teleparallel connection for the computation of the energy-momentum of the 
Schwarzschild solution. The generalization to the rotating Kerr
configurations will be discussed separately in the next section. 

We will consider several choices of the coframe. In order to show
how the covariant formulation works, we will compare our results
to the computations done in the pure tetrad formalism. The latter
gives, depending on the choice of the reference system, either
infinite or trivial answers. In contrast, the use of the covariant
teleparallel framework always yields the physically meaningful
result. 

\subsection{Schwarzschild metric: naive choice of a tetrad}\label{ES1}

In accordance with the spherical symmetry of the Schwarzschild
solution, we choose the spherical local coordinates, $(t, r,
\theta, \varphi)$. We start our analysis by using the diagonal coframe:
\begin{equation}
\vartheta^{\hat 0} = {\frac 1\alpha}\,cdt,\quad \vartheta^{\hat 1} 
= \alpha\,dr,\quad \vartheta^{\hat 2} = r\,d\theta,\quad 
\vartheta^{\hat 3} = r\sin\theta\,d\varphi,\label{cofS0}
\end{equation}
with $\alpha = \alpha(r)$. Actually, this class of metrics includes
not only Schwarzschild, but also Reissner-Nordstrom (with 
electric charge) and Kottler (with a cosmological term) metrics. 
The pure Schwarzschild arises when 
\begin{equation}
\alpha = \left(1 - {\frac {2m}r}\right)^{-{\frac 12}},
\end{equation}
with $m = GM/c^2$ ($G$ is Newtonian gravitational constant).

If we take tetrad (\ref{cofS0}), as well as the trivial 
Weitzenb\"ock connection $\Gamma_\alpha{}^\beta =0$, and substitute
them into (\ref{HEtilde}), we find 
\begin{equation}
\tilde{H}_{\hat 0} = {\frac \alpha\kappa}\,\cos\theta\,dr\wedge d\varphi 
-{\frac {2r}{\kappa\alpha}}\,\sin\theta\,d\theta\wedge d\varphi.\label{Hs}
\end{equation}
In particular, if we compute the total energy at a fixed time 
in the 3-space with a spatial boundary 2-dimensional surface 
$\partial S = \left\{r = R,\theta,\varphi\right\}$, with 
$R\rightarrow\infty$, we obtain 
\begin{equation}
\tilde{P}_{\hat{0}} = \int_{\partial S}\tilde{H}_{\hat{0}} = 
-\,{\frac {2R}{\kappa\alpha}}\int_{\partial S}\sin\theta\,
d\theta\wedge d\varphi = -\,{\frac {8\pi R}{\kappa\alpha}},
\end{equation}
which diverges in the limit of $R\rightarrow\infty$ (note that
$\alpha\rightarrow 1$ when the radius goes to infinity).

The physical reason that underlies such a result is obvious---the
energy-momentum current and the superpotential contain an infinite
contribution of the inertial effects that are present due to the
inconvenient choice of the reference system. We have demonstrated
in \cite{conserved} how to regularize this result by subtracting
the unphysical contribution with the help of the suitable choice
of the flat background connection. 
Here we use a different regularization framework which is based on
the covariance property. Namely, let us associate with the tetrad
(\ref{cofS0}) a nontrivial teleparallel connection 
\begin{equation}\label{flatGam}
\Gamma_{\hat{1}}{}^{\hat{2}} = d\theta,\qquad \Gamma_{\hat{1}}
{}^{\hat{3}} = \sin\theta d\varphi,\qquad \Gamma_{\hat{2}}{}^{\hat{3}} 
= \cos\theta d\varphi.
\end{equation}
The curvature obviously vanishes for this connection, but torsion
$T^\alpha = d\vartheta^\alpha +\Gamma_\beta{}^\alpha\wedge\vartheta^\beta$
is nontrivial.
Substituting into (\ref{H0K}), we then find
\begin{equation}
H_{\widehat 0} = {\frac {2r(1-1/\alpha)}{\kappa}}
\,\sin\theta\,d\theta\wedge d\varphi.\label{Hsch1}
\end{equation}
The integral over the spatial boundary yields
\begin{equation}
P_{\hat{0}} = \int_{\partial S}\,H_{\hat{0}} = M. 
\end{equation}

It is worthwhile to note that the account for the teleparallel 
connection removed the divergence and automatically produced the 
physical result. Another remark is in order about the choice of the
connection (\ref{flatGam}). As one can immediately see, this is the
same flat connection that was earlier used in \cite{conserved} to
subtract the inertial effects. This is not a mere coincidence. In
Ref.~\cite{conserved} we have demonstrated the possibility of reinterpreting
the background flat connection as a Weitzenb\"ock connection in the
metric-affine approach to the translational gauge gravity theory.

\subsection{Schwarzschild metric: proper tetrad}\label{ES2}

Instead of the inconvenient reference system of the previous section,
we now choose a coframe that represents what we will call 
a {\it proper tetrad}. The definition will be provided later. Since
a coframe is a basis of the cotangent space, it can be expanded 
with respect to a different basis. In other words, a new coframe 
can be always obtained from the old one with the help of the local
Lorentz rotation. 

We begin with an observation that the role of the teleparallel 
connection, that we used above, was to to remove (or to separate)
the inertial contribution from the truly gravitational one. By 
definition, since the teleparallel curvature is zero, the connection
is a ``pure gauge'', that is
\begin{equation}\label{gamW}
\Gamma_\alpha{}^\beta=(\Lambda^{-1})^\beta{}_\gamma d\Lambda^\gamma{}_\alpha , 
\end{equation}
The Weitzenb\"ock connection always has the form (\ref{gamW}). Since the 
Lorentz matrix $\Lambda^\alpha{}_\beta$ has to do with transformations among 
different frames, $\Gamma_\alpha{}^\beta$ turns out to describe inertial 
properties of a tetrad. 
In particular, it is easy to see that (\ref{flatGam}) is of the form
(\ref{gamW}), with the Lorentz matrix given explicitly by
\begin{equation}
\Lambda^\alpha{}_\beta = \left(\begin{array}{cccc}
1 & 0 & 0 & 0\\ 0 & \cos\varphi\sin\theta & \cos\varphi\cos\theta & 
-\sin\varphi \\
0 & -\cos\theta & \sin\theta & 0\\ 0 & \sin\varphi\sin\theta & \sin\varphi
\cos\theta & \cos\varphi\end{array}\right).\label{LR}
\end{equation}

Now, we are in a position to construct the proper tetrad. Qualitatively,
it is clear what we need to do. The reference system described in the 
previous section is ``spoiled'' by the presence of the inertial effects, 
so that the teleparallel connection was required for the regularization 
of the energy-momentum. These inertial effects are encoded in the Lorentz
matrix (\ref{LR}). Accordingly, in order to improve the situation, we need 
to go to a new reference system by performing the local Lorentz rotation
that removes the drawbacks mentioned. In technical terms, we define a
new tetrad ${\stackrel 0 \vartheta}{}^\alpha$ with the help of the  
Lorentz transformation 
\begin{equation}
{\stackrel 0 \vartheta}{}^\alpha = \Lambda^\alpha{}_\beta
\,\vartheta^\beta\label{cofP}
\end{equation}
with the rotation matrix (\ref{LR}). From (\ref{cofcontrans}) we can 
easily verify that the corresponding teleparallel connection, that is
associated to the new tetrad, is trivial:
\begin{equation}
{\stackrel 0 \Gamma}{}_\alpha{}^\beta = (\Lambda^{-1})^\mu{}_{\alpha}
\Gamma_\mu{}^\nu \Lambda^\beta{}_\nu + \Lambda^\beta{}_\gamma d(\Lambda^{-1}
)^\gamma{}_\alpha = 0.\label{gamP}
\end{equation}

In addition, the new coframe ${\stackrel 0 \vartheta}{}^\alpha$ has 
another important property: Let us ``switch off" the gravitational 
effects. Technically, one can do it by putting equal zero the essential 
gravitational parameters that describe a given configuration; in this 
case $m = 0$. After doing this, we discover that such a ``gravity 
switched-off'' tetrad becomes {\it holonomic}: 
\begin{equation}\label{FP}
{\stackrel 0 F}{}^\alpha =d{}{\stackrel 0 \vartheta}{}^\alpha{}\vline_{m=0} = 0.
\end{equation}
Actually, it is easy to see that the tetrad (\ref{cofP}) describes the 
Cartesian coordinate system when $m$ vanishes. This means, in physical 
terms, that {\it this frame does not include inertial effects}. 
This is the definition of {\it proper tetrad}: it is a 
coframe that describes a reference system whose anholonomy has to do 
with gravitation only, not with inertial effects. It corresponds, in 
this sense, to the inertial frames of special relativity, and it
reduces to a Cartesian frame in the absence of gravitation. 

Let us calculate the energy-momentum. 
Using (\ref{cofP}) and (\ref{gamP}) in (\ref{Ha}), we find
\begin{equation}
H_{\widehat 0} = \widetilde{H}_{\widehat 0} = {\frac {2r(1-1/\alpha)}{\kappa}}
\,\sin\theta\,d\theta\wedge d\varphi.
\end{equation}
The total energy is found to be finite: $P_{\widehat 0} =\int H_{\widehat 0} = M$.
The regularization is not needed! The energy-momentum is regular for the
proper tetrad, which is consistent with the fact that the inertial 
effects, that are responsible for the bad behavior of the energy 
and momentum, are absent in the proper reference system. 

\subsection{Schwarzschild metric: freely falling tetrad}\label{ES3}

Since Einstein with his famous thought experiments with an elevator, 
we know that gravity can be locally imitated by inertial effects, 
or alternatively, gravitational effects can be locally eliminated by
using an appropriate non-inertial reference system. A freely falling
elevator is an example. This fact constitutes the contents of the 
strong equivalence principle which underlies Einstein's gravity theory. 

A natural question then arises: What about the energy of the gravitational
field? What happens to the energy when we ``eliminate'' gravity by going
to a non-inertial system? One possible answer was recently proposed in
\cite{Maluf} in the framework of the pure tetrad (non-covariant) formulation. 
Here we will discuss the result of \cite{Maluf} and in the next section 
we will reanalyze the same question in the covariant formulation. 

We start again from the diagonal tetrad (\ref{cofS0}), and construct 
a new coframe with the help of the Lorentz transformation ${\stackrel f
\vartheta}{}^\alpha = \Lambda^\alpha{}_\gamma \Lambda'^\gamma{}_\delta
\,\vartheta^\delta$, where $\Lambda^\alpha{}_\gamma$ is given by 
(\ref{LR}), and 
\begin{equation}
\Lambda'^\gamma{}_\delta = \left(\begin{array}{cccc}
\alpha & \alpha\beta & 0 & 0\\ \alpha\beta & \alpha & 0 & 0 \\
0 & 0 & 1 & 0\\ 0 & 0 & 0 & 1\end{array}\right),\label{L}
\end{equation}
with $\beta = \sqrt{1 - \alpha^{-2}}$ (and hence $\alpha = 1/\sqrt{1
- \beta^2}$). Specifically for the Schwarzschild metric, we have
\begin{equation}
\beta = \sqrt{\frac {2m}r},\qquad \alpha = \left(1 - {\frac {2m}r}
\right)^{-{\frac 12}}.
\end{equation}

A direct computation shows that the superpotential is identically zero
for this tetrad: 
\begin{equation}
\widetilde{H}_{\widehat 0} = 0.\label{Hzero} 
\end{equation}
This was demonstrated by Maluf et al \cite{Maluf}. In physical terms, 
such a tetrad describes a reference frame that is freely falling along 
the radial coordinate onto the attracting source. Gravity is 
``eliminated'' in such a non-inertial system, and at the first sight 
the trivial result (\ref{Hzero}) seems to be a natural outcome. However, 
it is interesting to ask whether we still can calculate the total energy 
of the source, after all, the latter did not disappear physically
in the new reference system. We will come back to this question in
the next section, whereas here our aim is to analyze the origin of
the trivial result (\ref{Hzero}). 

In order to do this, we will make use of the object that is often called 
the ``generalized acceleration'' \cite{Maluf,Mashhoon}. Let us take the
frame $e_\alpha = h_\alpha^i\partial_i$, dual to the coframe
$\vartheta^\alpha = h^\alpha_idx^i$. The zeroth
leg of the frame
\begin{equation}
e_{\hat 0} = u,
\end{equation}
is usually interpreted as the 4-velocity of an observer, and the total
frame then represents a comoving reference system of an observer. 
The ``generalized acceleration'' object is defined by
\begin{equation}
\Phi_\alpha{}^\beta := h^\beta_i\,{\frac {\tilde{D}h^i_\alpha}{ds}},\label{fab}
\end{equation}
where $\tilde{D}h^i_\alpha = dh^i_\alpha + \tilde{\Gamma}_j{}^ih^j_\alpha$
is a covariant derivative with respect to the Riemannian (Christoffel) 
connection. It acts on the vector index $^i$, whereas the local tetrad 
index is just a label of the four legs of the frame. 

By definition, this object is not a tensor, which is a well known fact. 
Indeed, by using the standard relation for the components of the connection
in different frames, we easily find $h^\beta_i\tilde{D}h^i_\alpha = 
h^\beta_idh^i_\alpha + h^\beta_i\tilde{\Gamma}_j{}^ih^j_\alpha =
\tilde{\Gamma}_\alpha{}^\beta$. Consequently, we have explicitly
\begin{equation}
\Phi_\alpha{}^\beta = u\rfloor \tilde{\Gamma}_\alpha{}^\beta =
\tilde{\Gamma}_{{\hat 0}\alpha}{}^\beta.\label{fab2}
\end{equation}
In other words, the components of the ``generalized acceleration'' 
object coincide with some components of the Riemannian connection.

One can say that the condition 
\begin{equation}
\Phi_\alpha{}^\beta = 0\label{phi0}
\end{equation} 
defines a kind of inertial reference system. It is easy to see that 
$\Phi_{\hat 0}{}^\beta = a^\beta = h^\beta_i\,a^i$ with $a^i = u^k
\tilde{\nabla}_ku^i$ the acceleration. Accordingly, when $\Phi_\alpha
{}^\beta = 0$, the observer is ``freely falling'' without acceleration, 
and vanishing of the spatial components $\Phi_a{}^b$ ($a,b,\dots = 1,2,3$) 
means that the comoving triad of an observer is not rotating. 

Suppose that we have a reference system (a tetrad) with the property 
(\ref{phi0}). Is this a sufficient condition for the energy to vanish? 
To find this out, we recall that in the pure tetrad formulation the 
energy is calculated with the help of the superpotential (\ref{Ht}). 
We can straightforwardly see how the latter is related to the ``generalized 
acceleration'' object. We expand the connection 1-form with respect to 
the coframe basis, $\tilde{\Gamma}^{\beta\gamma} =
\vartheta^\lambda\tilde{\Gamma}_\lambda{}^{\beta\gamma}$, and then find
$\tilde{\Gamma}^{\beta\gamma}\wedge\eta_{\alpha\beta\gamma} = 
\tilde{\Gamma}_\alpha{}^{\beta\gamma}\eta_{\beta\gamma} 
+ 2\tilde{\Gamma}_\lambda{}^{\beta\lambda}\eta_{\alpha\beta}$. 
Accordingly, the zeroth-component of the superpotential (\ref{Ht}) reads
\begin{eqnarray}
\tilde{H}_{\hat 0} &=& {\frac 1 {2\kappa}}\left(\tilde{\Gamma}_{\hat 0}
{}^{\beta\gamma}\eta_{\beta\gamma} + 2\tilde{\Gamma}_\lambda{}^{\beta\lambda}
\eta_{{\hat 0}\beta}\right)\nonumber\\
&=& {\frac 1 {2\kappa}}\left(\Phi^{ab}\,\eta_{ab}
+ 2\tilde{\Gamma}_b{}^{ab}\,\eta_{{\hat 0}a}\right).\label{H0}
\end{eqnarray}

Thus, we see that, contrary to the assumption of \cite{Maluf}, 
the condition (\ref{phi0}) of the vanishing ``generalized acceleration''
is not responsible for the ``zero-energy'' result (\ref{Hzero}). Instead, 
we find from (\ref{H0}) that the vanishing rotation is indeed needed:
$\Phi^{ab} = 0$, $a,b = 1,2,3$. However, the absence of acceleration
$a^\beta$ is not necessary. In addition, however, one needs a rather 
curious condition for the 3D trace of the connection $\tilde{\Gamma}_b
{}^{ab} =0$. As a matter of fact, the 6 conditions
\begin{equation}
\Phi^{ab} = 0,\qquad \tilde{\Gamma}_b{}^{ab} = 0,\label{cond}
\end{equation}
can be used to fix the choice of the tetrad, thus eliminating the 
freedom of the 6-parameter local Lorentz transformations. It is 
unclear though if this gauge is useful in practice.
One can prove by a direct inspection that indeed the condition 
(\ref{cond}) is fulfilled for the freely falling tetrad ${\stackrel f
\vartheta}{}^\alpha$. As we will demonstrate later, a similar freely 
falling tetrad can be constructed also for the Kerr solution.

\subsection{Schwarzschild metric: free fall in the covariant formulation}
\label{ES4}

Let us now reanalyze the same question in the covariant formulation. 
We expect that taking appropriately into account the teleparallel 
connection (that is responsible for the inertial effects, as we 
already know), it will become possible to clear the gravitational
energy of the contributions coming from the non-inertial dynamics of
the reference system. Indeed, this can be perfectly confirmed by
explicit computations as follows. 

Using (\ref{cofcontrans}), we straightforwardly find the teleparallel
connection associated with the ``freely falling'' tetrad ${\stackrel f
\vartheta}{}^\alpha$. It reads 
\begin{equation}
{\stackrel f \Gamma}{}_\alpha{}^\beta=({\stackrel f \Lambda}{}^{-1})^\beta
{}_\gamma\,{}d{}\,{\stackrel f \Lambda}{}^\gamma{}_\alpha ,\label{gamF}
\end{equation}
where ${\stackrel f \Lambda}{}^\alpha{}_\beta =\Lambda^\alpha
{}_\gamma(\Lambda'{}^{-1})^\gamma{}_\delta(\Lambda^{-1})^\delta{}_\beta$.
The Weitzenb\"ock torsion for $({\stackrel f \vartheta}{}^\alpha, 
{\stackrel f \Gamma}{}_\alpha{}^\beta)$ has a rather complicated form,
but using it in (\ref{Ha}) we find for the superpotential
\begin{equation}
H_{\widehat 0} = {\frac {2r(\alpha - 1)}{\kappa}}
\,\sin\theta\,d\theta\wedge d\varphi.
\end{equation}
The total energy is thus $P_{\widehat 0} =\int H_{\widehat 0} = M$, as 
before. It is satisfactory to see that the final result is neither
infinity nor zero. The teleparallel connection (\ref{gamF}) 
has automatically regularized the situation. The contribution due to the 
non-inertial motion of the ``freely falling'' reference system (that
compensated the gravitational one in the purely tetrad formulation) is
now subtracted and the correct total energy of the source is recovered.

\section{Energy for the rotating Kerr solution}\label{KS}

Although the Schwarzschild solution is a special case of the Kerr 
solution, we analyze these cases separately. The reason is that the 
Kerr metric is essentially more complicated and its study requires 
some specific techniques, which are not needed in the Schwarzschild 
case. Moreover, the final formulas are usually very nontrivial for
the Kerr configuration and one needs to make various approximations
(taking the limit of infinite radius, for example), whereas in the
previous section it was possible to give the exact expressions. 

In our discussion we use a spherical type local coordinate system 
$(t,r,\theta,\varphi)$ that is known also as the Boyer-Lindquist 
coordinate system. 

\subsection{Kerr metric: a naive tetrad}\label{KS1}

We will follow closely the scheme outlined earlier for the 
Schwarzschild metric, and choose the tetrad in the form
\begin{equation}
\vartheta^{\hat 0} = {\frac {\sqrt{\Sigma\Delta}}{\cal A}}\,cdt,\quad 
\vartheta^{\hat 1} = \sqrt{\frac \Sigma\Delta}\,dr,\quad 
\vartheta^{\hat 2} = \sqrt{\Sigma}\,d\theta,\quad 
\vartheta^{\hat 3} = {\frac {\sin\theta}{\sqrt{\Sigma}}}\left(
{\cal A}\,d\varphi - {\frac {2amr}{\cal A}}\,dt\right).\label{cofK0}
\end{equation}
Here the functions and constants are defined by
\begin{eqnarray}
\Delta&:=& r^2 + a^2  - 2mr,\\
\Sigma&:=& r^2 + a^2\cos^2\theta,\\
m &:=& \frac{GM}{c^2},\\
{\cal A}^2 &=&  \Delta\Sigma + 2mr(r^2 + a^2) \equiv (r^2 + a^2)^2 
- a^2\sin^2\theta\,\Delta.
\end{eqnarray}
As we can immediately check, this tetrad reduces to the diagonal 
coframe (\ref{cofS0}) when we put the rotation parameter equal to zero:
$a = 0$. After noticing this direct relation to the diagonal tetrad
of the Schwarzschild solution, we can expect similar problems for
the computation of the energy-momentum. As a matter of fact, this is
indeed the case.

For the tetrad (\ref{cofK0}), accompanied by the trivial 
Weitzenb\"ock connection $\Gamma_\alpha{}^\beta =0$, we find the 
superpotential 
\begin{eqnarray}
\tilde{H}_{\hat 0} &=& {\frac {am}{\kappa{\cal A}\sqrt{\sigma\Delta}}}
\,cdt\wedge\left(2r\cos\theta\,dr - \Delta\sin\theta\,d\theta\right)
+ {\frac {{\cal A}\cos\theta}{\kappa\sqrt{\sigma\Delta}}}
\,dr\wedge d\varphi\nonumber\\ 
&& -\,{\frac {\sqrt{\Delta}\left[2r(r^2 + a^2) + (m - r)a^2\sin^2\theta\right]}
{\kappa{\cal A}\sqrt{\Sigma}}}\,\sin\theta\,d\theta\wedge d\varphi.\label{HKs}
\end{eqnarray}
The last term has the leading behavior $\sim 2r$, just like the last
term in (\ref{Hs}), and thus the total energy (calculated as the integral 
over the sphere of infinite radius) is divergent. 

As a check, we can straightforwardly verify that the tetrad (\ref{cofK0})
is not holonomic as such, and the tetrad that is obtained from it by
``switching off'' gravity (putting $m = 0$ and $a = 0$) is anholonomic
too. This means that the inertial effects are again ``spoiling'' the
picture. 

The regularization is needed and as before this is achieved with the help 
of the same teleparallel connection (\ref{flatGam}).
Substituting now the pair 
$(\vartheta^\alpha, \Gamma_\alpha{}^\beta)$, where the tetrad is given by 
(\ref{cofK0}) and the connection by (\ref{flatGam}), into (\ref{H0K}), we find
\begin{equation}
H_{\widehat 0} = \left({\frac {2m}{\kappa}} + \dots
\right)\sin\theta\,d\theta\wedge d\varphi + \dots.\label{Hkerr1}
\end{equation}
The resulting expressions are rather complicated for the Kerr metric,
so from now on we will display only the leading terms, whereas the 
terms that are proportional to $1/r^n$, $n\geq 1$ will be denoted 
by the dots. The integral over the spatial boundary then yields
\begin{equation}
P_{\hat{0}} = \int_{\partial S}\,H_{\hat{0}} = M. 
\end{equation}

\subsection{Kerr metric: proper tetrad}\label{KS2}

The proper tetrad is constructed along the same lines as we did in the
previous section. Since the regularizing teleparallel connection has 
the form (\ref{gamW}), we define the proper tetrad as the coframe 
${\stackrel 0 \vartheta}{}^\alpha$ that is obtained from (\ref{cofK0}) 
with the help of the Lorentz transformation (\ref{LR}). It is easy to 
see that the resulting coframe indeed has the required properties: 
(i) the teleparallel connection is zero (\ref{gamP}), (ii) the tetrad 
${\stackrel 0 \vartheta}{}^\alpha$ becomes holonomic (\ref{FP}) when 
the gravity is ``switched off'' (for $m=0$, $a=0$). 

The computation of the energy and momentum for the proper Kerr tetrad
is straightforward. The result reads (again giving the leading terms
only) as follows:
\begin{equation}
H_{\widehat 0} = \widetilde{H}_{\widehat 0} = {\frac {2m + (m^2 + a^2
- {\frac 12}a^2\sin^2\theta)/r + {\cal O}(1/r^2)}{\kappa}}\,\sin\theta
\,d\theta\wedge d\varphi + \dots.
\end{equation}
The total energy is finite, $P_{\widehat 0} =\int H_{\widehat 0} = M$.
Thus again we prove that for the proper tetrad one does not need a
regularization, the result for the total energy-momentum is automatically
finite and has the correct value.

\subsection{Kerr metric: freely falling tetrad}\label{KS3}

Here we study the possibility to find a non-inertial reference system 
in which gravitation is ``eliminated'' by the inertia. Such a generalization
of a freely falling tetrad from the Schwarzschild to the Kerr case can
be indeed constructed. 

As a first step, let us make a local Lorentz transformation 
\begin{equation}
{\stackrel d \vartheta}{}^\alpha = {\stackrel d \Lambda}{}^\alpha{}_\beta
\,\vartheta^\beta,\label{cofD}
\end{equation}
where ${\stackrel d \Lambda}{}^\alpha{}_\beta =(\Lambda_2){}^\alpha{}_\gamma
(\Lambda_1){}^\gamma{}_\beta$, with
\begin{equation}
(\Lambda_1){}^\alpha{}_\beta = \left(\begin{array}{cccc}
{\cal A}/\sqrt{\Delta\Sigma} & \sqrt{2mr(r^2 + a^2)}/\sqrt{\Delta\Sigma} 
& 0 & 0\\ \sqrt{2mr(r^2 + a^2)}/\sqrt{\Delta\Sigma} & {\cal A}/\sqrt{\Delta
\Sigma} 
& 0 & 0 \\ 0 & 0 & 1 & 0\\ 0 & 0 & 0 & 1\end{array}\right),\label{L1}
\end{equation}
\begin{equation}
(\Lambda_2){}^\alpha{}_\beta = \left(\begin{array}{cccc}
1 & 0 & 0 & 0\\ 0 & \sqrt{\Sigma(r^2 + a^2)}/{\cal A} & 0 
& -a\sin\theta\sqrt{2mr}/{\cal A} \\ 0 & 0 & 1 & 0\\ 
0 & a\sin\theta\sqrt{2mr}/{\cal A} & 0 & \sqrt{\Sigma(r^2 + a^2)}/{\cal A}
\end{array}\right).\label{L2}
\end{equation}
This brings us from the original tetrad (\ref{cofK0}) to the coframe
\begin{eqnarray}
{\stackrel d \vartheta}{}^{\hat 0} &=& cdt 
+ {\frac {\sqrt{2mr(r^2 + a^2)}}{\Delta}}\,dr,\label{coft0} \\
{\stackrel d \vartheta}{}^{\hat 1} &=& \sqrt{\frac {2mr}{\Sigma}}
\left(cdt - a\sin^2\theta\,d\varphi\right) 
+ {\frac {\sqrt{\Sigma(r^2 + a^2)}}{\Delta}}\,dr,\label{coft1}\\
{\stackrel d \vartheta}{}^{\hat 2} &=& \sqrt{\Sigma}\, d\theta,\label{coft2}\\
{\stackrel d \vartheta}{}^{\hat 3} &=& \sin\theta\left(\sqrt{r^2 + a^2}
\,d\varphi + {\frac {a\sqrt{2mr}}{\Delta}}\,dr\right).\label{coft3}
\end{eqnarray}
We will call this new coframe a Doran tetrad because 
(\ref{coft0})-(\ref{coft3}) is closely related to an alternative 
representation of the Kerr metric given by Doran in \cite{Doran}. We can 
simplify the above formulas by making the coordinate transformations
\begin{eqnarray}
cdt_d &=& cdt + {\frac {\sqrt{2mr(r^2 + a^2)}}{\Delta}}\,dr,\\ \label{phiD}
d\varphi_d &=& d\varphi + {\frac a\Delta}\sqrt{\frac {2mr}{r^2 + a^2}}\,dr.
\end{eqnarray}
In the new ``Doran coordinates'' $(t_d, r, \theta, \varphi_d)$, the Kerr metric
will reduce to the form described in \cite{Doran}. Indeed, the coframe 
(\ref{coft0})-(\ref{coft3}) in the new coordinates reads
\begin{eqnarray}
{\stackrel d \vartheta}{}^{\hat 0} &=& cdt_d,\label{cofd0} \\
{\stackrel d \vartheta}{}^{\hat 1} &=& \sqrt{\frac {2mr}{\Sigma}}
\left(cdt_d - a\sin^2\theta\,d\varphi_d\right) 
+ \sqrt{\frac \Sigma {r^2 + a^2}}\,dr,\label{cofd1}\\
{\stackrel d \vartheta}{}^{\hat 2} &=& \sqrt{\Sigma}\, d\theta,\label{cofd2} \\
{\stackrel d \vartheta}{}^{\hat 3} &=& \sin\theta\sqrt{r^2 + a^2}
\,d\varphi_d.\label{cofd3}
\end{eqnarray}
This exactly reproduces the line element of Doran, see eq. (18) in \cite{Doran}.

One can verify that the zeroth leg of the dual frame
\begin{equation}
u = {\stackrel d e}{}_{\hat 0} = {\frac {\cal A} {c\Sigma\Delta}}\partial_t 
- {\frac {\sqrt{2mr(r^2 + a^2)}}{\Sigma}}\,\partial_r 
+ {\frac {2amr}{\Sigma\Delta}}\,\partial_\varphi = {\frac 1c}\partial_{t_d} 
- {\frac {\sqrt{2mr(r^2 + a^2)}}{\Sigma}}\,\partial_r
\end{equation}
is a geodesic vector field, the acceleration is zero $u^i\nabla_i u^j = 0$,
or $\Phi_{\hat 0}{}^a = 0$. However, this frame has a nontrivial rotation, 
namely
\begin{equation}
\Phi^{{\hat 1}{\hat 2}} = a^2\sin\theta\cos\theta
\,{\frac {\sqrt{2mr}}{\Sigma^2}}. 
\end{equation}
This reference system thus does not satisfy the ``compensation 
conditions'' (\ref{cond}). 

However, we can improve the situation if we make an additional Lorentz 
transformation
\begin{equation}
{\stackrel f \vartheta}{}^\alpha = (\Lambda_4){}^\alpha{}_\gamma
(\Lambda_3){}^\gamma{}_\beta\,{\stackrel d \vartheta}{}^\beta,\label{cofKF}
\end{equation}
where the matrices
\begin{equation}
(\Lambda_3){}^\alpha{}_\beta = \left(\begin{array}{cccc}
1 & 0 & 0 & 0\\ 0 & \cos\theta\,\sqrt{(r^2 + a^2)/\Sigma} & 
-\sin\theta\,r/\sqrt{\Sigma} & 0 \\ 0 & \sin\theta\,r/\sqrt{\Sigma} & 
\cos\theta\,\sqrt{(r^2 + a^2)/\Sigma} & 0\\ 
0 & 0 & 0 & 1\end{array}\right),\label{L4}
\end{equation}
and 
\begin{equation}
(\Lambda_4){}^\alpha{}_\beta = \left(\begin{array}{cccc}
1 & 0 & 0 & 0\\ 0 & 0 & \cos\varphi_d & -\sin\varphi_d \\ 
0 & 0 & \sin\varphi_d & \cos\varphi_d \\ 
0 & 1 & 0 & 0\end{array}\right).\label{L5}
\end{equation}
Note that the transformation (\ref{L4}) is the same in both coordinate 
systems, in the original $(t, r, \theta, \varphi)$ and in the Doran 
coordinates $(t_d, r, \theta, \varphi_d)$. However the transformation 
(\ref{L5}) refers to the Doran coordinate system. Of course one can apply 
it also in the original coordinates, but keep in mind that $\varphi_d$ 
is a function defined by the integral from (\ref{phiD}). 

One can verify that the ``generalized acceleration'' object vanishes   
for the final coframe, $\Phi_\alpha{}^\beta = 0$. This means that indeed
the resulting tetrad does not have acceleration and rotation, i.e., 
it describes a ``freely falling'' reference system. However, still 
$\tilde{\Gamma}_b{}^{ab}\neq 0$ for this frame. Namely,
\begin{eqnarray}
\tilde{\Gamma}_b{}^{\hat{1}b} &=& {\frac {a\sin\varphi_d\sin\theta}{\Sigma}}
\,\sqrt{\frac {2m}{r}},\\
\tilde{\Gamma}_b{}^{\hat{2}b} &=& {\frac {-\,a\cos\varphi_d\sin\theta}{\Sigma}}
\,\sqrt{\frac {2m}{r}}
\end{eqnarray}

As a result, the conditions (\ref{cond}) are not satisfied for this
frame, and the tetrad energy-momentum does not vanish, in contrast to 
(\ref{Hzero}). Instead, the direct computation of the $\widetilde{H}_\alpha$ 
for the final coframe ${\stackrel f \vartheta}{}^\alpha$ yields:
\begin{eqnarray}
\widetilde{H}_{\hat 0} &=& a\sin\theta\,\left[{\frac {2m}{\Sigma}}
\left(cdt_d - a\sin2\theta\,d\varphi_d\right) + \sqrt{\frac {2m}
{r(r^2 + a^2)}}\,dr\right]\wedge d\theta,\label{H0t}\\
\widetilde{H}_{\hat 1} &=& {\frac {\sin^2\theta}\Sigma}\Big[\cos\varphi_d
\,\sqrt{2mr}\left(a^2 - 2r^2(r^2 + a^2)/\Sigma\right) \nonumber\\
&&\qquad +\,\sin\varphi_d\,am\sqrt{r^2 + a^2}\Big] 
d\varphi_d\wedge d\theta + \dots,\label{H1t}\\
\widetilde{H}_{\hat 2} &=& {\frac {\sin2\theta}\Sigma}\big[\sin\varphi_d
\,\sqrt{2mr}\left(a^2 - 2r^2(r^2 + a^2)/\Sigma\right) \nonumber\\
&&\qquad -\,\cos\varphi_d\,am\sqrt{r^2 + a^2}\big] 
d\varphi_d\wedge d\theta + \dots,\label{H2t}\\
\widetilde{H}_{\hat 3} &=& -\,{\frac {\sin\theta\cos\theta}{\Sigma^2}}
\,\sqrt{m}\left[2r(r^2 + a^2)\right]^{3/2}\,d\varphi_d \wedge d\theta 
+ \dots\,.\label{H3t}
\end{eqnarray}
The expression (\ref{H0t}) is exact, whereas in (\ref{H1t})-(\ref{H3t}) the
dots denote other terms which are irrelevant for the computation of the 
total conserved quantities in a sphere of an arbitrary radius. It is easy
to see that integration over the spherical angles ($\int_0^{2\pi}d\varphi_d
\int_0^\pi d\theta$, note that $d\varphi_d = d\varphi$ on a sphere) yields 
zero for the components (\ref{H1t})-(\ref{H3t}). 

Thus, the local energy density is nontrivial despite the fact that the 
tetrad is non-accelerating and non-rotating. Nevertheless, since in the limit
of the large radius we have the leading behavior $\widetilde{H}_{\hat 0} 
\cong r^{-2}$, the {\it total energy} is obviously zero for this coframe:
\begin{equation}
P_{\widehat 0} =\int \widetilde{H}_{\widehat 0} = 0.\label{Pzero}
\end{equation}
This result again originates from the contribution of the essentially 
non-inertial behavior of the reference system. 

\subsection{Kerr metric: covariant treatment of the freely falling  frame}
\label{KS4}

As in the Schwarzschild case, the situation can be improved if the 
reanalyze the same problem in the covariant teleparallel framework. 
In this case we start with the original coframe (\ref{cofK0}) and
the corresponding regularizing teleparallel connection (\ref{flatGam}),
and construct for the final coframe ${\stackrel f \vartheta}{}^\alpha$
the corresponding Weitzenb\"ock connection from the transformation 
law (\ref{cofcontrans}):
\begin{equation}\label{gamKF}
{\stackrel f \Gamma}{}_\alpha{}^{\beta} = ({\stackrel f \Lambda}{}^{-1})^\mu
{}_{\alpha}\Gamma_\mu{}^\nu {\stackrel f \Lambda}{}^\beta{}_{\nu} 
+ {\stackrel f \Lambda}{}^\beta{}_{\gamma}d({\stackrel f 
\Lambda}{}^{-1})^\gamma{}_{\alpha},
\end{equation}
where ${\stackrel f \Lambda}{}^\alpha{}_\beta =
(\Lambda_4){}^\alpha{}_\mu(\Lambda_3){}^\mu{}_\nu
(\Lambda_2){}^\nu{}_\lambda(\Lambda_1){}^\lambda{}_\beta$. 

The Weitzenb\"ock torsion for the final pair of fields 
$({\stackrel f \vartheta}{}^\alpha, {\stackrel f \Gamma}{}_\alpha
{}^\beta)$ is much more complicated than that of the Schwarzschild
metric. Nevertheless, it is straightforward to substitute it in 
(\ref{H0K}), (\ref{Ha}) and to get the superpotential
\begin{equation}
H_{\widehat 0} = {\frac {2m + (3m^2 + a^2
- {\frac 32}a^2\sin^2\theta)/r + {\cal O}(1/r^2)}{\kappa}}\,\sin\theta
\,d\theta\wedge d\varphi + \dots.
\end{equation}
The total energy is thus $P_{\widehat 0} =\int H_{\widehat 0} = M$, as 
before. We see again that the end result is neither infinity 
nor zero. The teleparallel connection (\ref{gamKF}) 
has automatically regularized the energy for the Kerr solution 
in the same way it worked for the Schwarzschild case. Namely, the 
contribution due to the non-inertial motion of the generalized ``freely 
falling'' reference system (in which gravity is locally eliminated) is
again correctly subtracted and the physically meaningful value for the
total energy of the source is again recovered.

\section{Discussion and conclusion}\label{DC}

Although equivalent to general relativity, teleparallel gravity has several 
conceptual differences with respect to general relativity. One of these 
differences is that the Weitzenb\"ock connection represents only inertial 
effects related to the frame. As a consequence of this property, one can
separate gravitation from inertial effects. It becomes then possible to
write down a tensorial expression for the energy-momentum density of 
gravity. Due to the fact that the frame-related inertial contribution to
the conserved quantities is always properly subtracted by the
Weitzenb\"ock connection, the covariant teleparallel approach naturally
yields regularized solutions for the energy and momentum.

As a test of the regularizing property of teleparallelism, we have 
considered in this paper two concrete examples: the Schwarzschild and 
Kerr solutions. For these two important cases, we have computed the total 
energy for different frames, and have shown that the covariant teleparallel 
approach always yields the physically correct result. We can 
thus say that the Weitzenb\"ock connection acts as a regularizing tool 
which separates the inertial energy-momentum density, leaving the tensorial, 
physical energy-momentum density of the system untouched.

\begin{acknowledgments}
The work of YNO was supported by FAPESP. The other authors thank 
FAPESP, CNPq and CAPES for partial financial support.
\end{acknowledgments}


\begin{thebibliography}{99}

\bibitem{Trautman}
A. Trautman, {\it Conservation laws in general relativity}, in: {\sl 
``Gravitation: An introduction to current research''}, L. Witten, ed. 
(John Wiley and Sons, New York, 1962) pp. 169-198. 

\bibitem{Faddeev}
L.D. Faddeev, {\it Problem of energy in Einstein's theory of gravity},
{\sl Sov. Phys. Usp.} {\bf 25} (1982) 130 [{\sl Usp. Fiz. Nauk}, {\bf 136}
(1982) 435-457 (in Russian)].

\bibitem{Szabados}
L.B. Szabados, {\it Quasi-local energy-momentum and angular momentum
in GR: A review article}, {\sl Living Rev. Rel.} {\bf 7} (2004) 4;
http://www.livingreviews.org/lrr-2004-4.

\bibitem{Blag}
M. Blagojevi\'c, {\it Gravitation and gauge symmetries} (Institute of
Physics: Bristol, 2002).

\bibitem{Nester}
J.M. Nester, {\it General pseudotensors and quasilocal quantities}, 
{\sl Class. Quantum Grav.} {\bf 21} (2004) S261-S280. 

\bibitem{Einstein05}
R. Aldrovandi, L.C.T. Guillen, J.G. Pereira, and K. H. Vu, 
{\it Bringing together gravity and the quanta}, contribution to 
the proceedings of the Albert Einstein Century International Conference, 
Paris, July 18-22, 2005 [gr-qc/0603122].

\bibitem{PeLu1}
R. Aldrovandi, Tiago Gribl Lucas, and J.G. Pereira, {\it Gravitational 
energy-momentum conservation in teleparallel gravity},
arXiv:0812.0034 [gr-qc].

\bibitem{Moller} 
C. M\o{}ller, {\it Conservation laws and absolute parallelism in general
relativity}, {\sl Mat. Fys. Skr. Dan. Vid. Selsk.} {\bf 1}, no. 10 (1961) 
1-50.

\bibitem{Pele}
C. Pellegrini and J. Plebanski, {\it Tetrad fields and gravitational fields},
{\sl Mat. Fys. Skr. Dan. Vid. Selsk.} {\bf 2}, no. 4 (1963) 1-39.

\bibitem{Kaempfer}
F.A. Kaempfer, {\it Vierbein field theory of gravity}, {\sl Phys. Rev.}
{\bf 165} (1968) 1420-1423. 

\bibitem{Rodicev}
V.I. Rodichev, {\it Theory of gravity in orthogonal rep\`ere} (Nauka,
Moscow, 1974) (in Russian).

\bibitem{Cho}
Y.M. Cho, {\it Einstein  Lagrangian  as  the  translational  Yang-Mills
Lagrangian}, {\sl Phys. Rev.} {\bf D14} (1976) 2521-2525.

\bibitem{HSh79}
K. Hayashi and T. Shirafuji, {\it New general relativity},
{\sl Phys. Rev.} {\bf D19} (1979) 3524-3553.

\bibitem{Meyer}
H. Meyer, {\it Moller's tetrad theory of gravitation as a special case
of Poincar\'e theory -- a coincidence?}  {\sl Gen. Relat. Grav.} {\bf 14}
(1982) 531-548.

\bibitem{HMMN95} 
F.W. Hehl, J.D. McCrea, E.W. Mielke, and Y. Ne'eman, {\it Metric-affine 
gauge theory of gravity: field equations, Noether identities, world spinors, 
and breaking of dilaton invariance}, {\sl Phys. Rep.} {\bf 258}  (1995) 1-171.

\bibitem{Gron}
F. Gronwald, {\it Metric-affine gauge theory of gravity. I: Fundamental
structure and field equations}, {\sl Int. J. Mod. Phys.} {\bf D6} (1997)
263-304.

\bibitem{Muench97}
U. Muench, {\it \"Uber teleparallele Gravitationstheorien}, Diploma Thesis, 
University of Cologne (1997).

\bibitem{dea}
V.C. de Andrade and J.G. Pereira, {\it Gravitational Lorentz force and 
the description of the gravitational interaction}, {\sl Phys. Rev.} 
{\bf D56} (1997) 4689-4695. 

\bibitem{Tres}
R. Tresguerres, {\it Translations and dynamics}, {\sl Int. J. Geom. Meth.
Mod. Phys.} {\bf 5} (2008) 905-946. 

\bibitem{telemag}
Yu.N. Obukhov and J.G. Pereira, {\it Metric-affine approach to teleparallel 
gravity}, {\sl Phys. Rev.} {\bf D67} (2003) 044016 (17 pages).

\bibitem{telemag2}
Yu.N. Obukhov and J.G. Pereira, {\it Lessons of spin and torsion: Reply to 
``Consistent coupling to Dirac fields in teleparallelism"}, {\sl Phys. Rev.}
{\bf D69} (2004) 128502 (3 pages).

\bibitem{noninv}
Yu.N. Obukhov, G.F. Rubilar and J.G. Pereira, {\it Conserved currents
in gravitational models with quasi-invariant Lagrangians: Application
to teleparallel gravity},
{\sl Phys. Rev.} {\bf D74} (2006) 104007 (2006) (10 pages).

\bibitem{PGrev}
Yu.N. Obukhov, {\it Poincar\'e gauge gravity: selected topics},
{\sl Int. J. Geom. Meth. Mod. Phys.} {\bf 3} (2006) 95-137. 

\bibitem{dAGP00}
V.C. de Andrade, L.C.T. Guillen, and J.G. Pereira, {\it Gravitational 
energy-momentum density in teleparallel gravity}, {\sl Phys. Rev. Lett.} 
{\bf 84} (2000) 4533-4536.

\bibitem{conserved}
Yu.N. Obukhov and G.F. Rubilar, {\it Covariance properties and regularization 
of conserved currents in tetrad gravity}, {\sl Phys. Rev.} {\bf D73} (2006) 
124017 (14 pages). 

\bibitem{Maluf}
J.W.~Maluf, F.F. Faria, and S.C. Ulhoa, {\it On reference frames in 
spacetime and gravitational energy in freely falling frames},
{\sl Class. Quantum Grav.} {\bf 24} (2007) 2743-2753.

\bibitem{Mashhoon}
B. Mashhoon, {\it Nonlocality of accelerated systems}, {\sl
Int. J. Mod. Phys.} {\bf D14} (2005) 171-179. 

\bibitem{Doran}
C. Doran, {\it New form of the Kerr solution},
{\sl Phys. Rev.} {\bf D61} (2000) 067503 (4 pages). 

\end{thebibliography}
\end{document}